# Summary of Information Theoretic Quantities

Robin A.A. Ince[1], Stefano Panzeri[1,2] and Simon R. Schultz[3]

[1] *Institute of Neuroscience and Psychology, 58 Hillhead Street, University of Glasgow, Glasgow G12 8QB, UK*

[2] *Center For Neuroscience and Cognitive Systems, Italian Institute of Technology, Corso Bettini 31 – 38068 Rovereto (Tn) Italy*

[3] *Department of Bioengineering, Imperial College London, South Kensington, London SW7 2AZ, UK*

7 pages 3443 words.

## Definition

Information theory is a practical and theoretical framework developed for the study of communication over noisy channels. Its probabilistic basis and capacity to relate statistical structure to function make it ideally suited for studying information flow in the nervous system. As a framework it has a number of useful properties: it provides a general measure sensitive to any relationship, not only linear effects; its quantities have meaningful units which in many cases allow direct comparison between different experiments; and it can be used to study how much information can be gained by observing neural responses in single experimental trials, rather than in averages over multiple trials. A variety of information theoretic quantities are in common use in neuroscience – including the Shannon entropy, Kullback-Leibler divergence, and mutual information. In this entry, we introduce and define these quantities. Further details on how these quantities can be estimated in practice are provided in the entry "Estimation of Information-Theoretic Quantities" and examples of application of these techniques in neuroscience can be found in the entry "Applications of Information-Theoretic Quantities in Neuroscience".

# Detailed Description

## Information theoretic quantities

### Entropy as a measure of uncertainty

Information theory derives from Shannon's theory of communication (Shannon, 1948; Shannon and Weaver, 1949). *Information*, as we use the word technically, is associated with the *resolution of uncertainty*. The underpinning theoretical concept in information theory is thus the measurement of uncertainty, for which Shannon derived a quantity called *entropy*, by analogy to statistical mechanics. Shannon proved that the only quantity suitable for measuring the uncertainty of a discrete random variable *X* is:

$$H(X) = -K \sum_x P(x) \log_2 P(x) \qquad (1)$$

where *X* can take a number of values *x* according to the probability distribution *P(x)*. The constant *K* we take to be one. When the logarithm is taken to base 2, the resulting units of the entropy *H(x)* are called *bits* (when the natural logarithm is used the term is *nats*).

Entropy can be thought of as a non-parametric way to measure the variability of a distribution. Spread out distributions (with high variability) will have high entropy since all potential outcomes have similar probabilities and so the outcome of any particular draw is very uncertain. On the other hand, concentrated distributions (with low variability) will have lower entropy, since some outcomes will have high probability, allowing a reasonable guess to be made about the outcome of any particular draw. In this respect, entropy can be thought of as a generalised form of variance; unlike variance, which is a 2$^{nd}$ order statistic and only meaningful for uni-modal distributions, entropy can give meaningful values for any form of distribution[1].

As an example, consider the roll of an unbiased 6 sided die performed under a cup. With no external knowledge about the roll, an observer would believe any of the numbers are equally likely – a uniform distribution over the 6 possible faces. From equation (1), the entropy of this distribution is $\log_2 6$. Now a third party peeks under the cup and tells our observer that the die is showing an even number. This knowledge reduces the uncertainty about the result of the roll, but by how much? After being told the die is showing an even number, the number of possibilities is reduced from 6 to 3, but each of the even numbers remain equally likely. The entropy of this distribution is $\log_2 3$. So the reduction in the observers uncertainty, measured as the difference in entropy, is $\log_2 6 - \log_2 3 = \log_2 2$, or 1 bit. We can quantify the knowledge imparted by the statement "the result is even" as 1 bit of information. This corresponds to a reduction in uncertainty of a factor of two (from 6 to 3 possible outcomes). For this example, in both the before and after situations all the possibilities were equally likely (uniform distributions) but the methodology can be applied to any possible distribution. Note that the uniform distribution is the maximum entropy distribution over a finite set. Any other distribution would have lower entropy since it is not possible to be less uncertain about a possible outcome than when all possibilities are equally likely – there is no structure to allow any sort of informed guess. For further applications of the concept of *maximum entropy* see the entry on "Estimating information-theoretic quantities".

---

[1] The differential entropy (the continuous analogue of the discrete entropy discussed here) of a Normal distribution is proportion to the logarithm of the variance.

The fact that *H(X)* should be maximised by a uniform distribution is one of three axioms Shannon started from in order to derive the form of Eq. (1). The others are that impossible events do not contribute to the uncertainty, and that the uncertainty from a combination of independent events should be the sum of the uncertainty of the constituent events. These three conditions necessarily lead to Eq. (1), although it can be reached via many other routes as well (Chakrabarti and Chakrabarty, 2005).

## Mutual Information

The example of a die roll motivates how a difference between entropies can quantify the information conveyed about a set of possible outcomes. In the case of two discrete random variables – here we consider *S*, representing a set of stimuli which are presented during an experiment, and *R*, a set of recorded responses – this is formalised in a quantity called the *mutual information*. This is a quantity measuring the dependence between the two random variables and which can be defined in terms of entropies in the following three equivalent forms:

$$\begin{aligned} I(R;S) &= H(S) - H(S\mid R) \\ &= H(R) - H(R\mid S) \\ &= H(R) + H(S) - H(R,S) \end{aligned} \qquad (2)$$

*H(R),H(S)* are the individual entropies of each random variable as discussed above, *H(R,S)* is the *joint entropy* of the two variables and *H(R|S), H(S|R)* are *conditional entropies*. The conditional entropy is defined as

$$H(X\mid Y) = \sum_y P(y) H(X\mid Y = y) \qquad (3)$$

Where *H(X|Y=y)* is defined as in Eq. (1), but with *P(x)* replaced by the conditional probability *P(x|y)*. The conditional entropy *H(X|Y)* represents the average entropy of *X* when the value of *Y* is known.

The three forms of Eq. (2) above each illustrate a particular interpretation of the mutual information. From the first form, we can see it quantifies the average reduction in uncertainty about which stimulus was presented based on observation of a response *R* answering the question "if we observe *R*, by how much does that reduce our uncertainty about the value of *S*?". Equivalently, the second form shows us that it similarly answers the question "if we observe *S*, by how much does that reduce our uncertainty about the value of *R*?". As the third form shows explicitly, mutual information is symmetric in its arguments. The third form also shows that information is the difference in entropy between a model in which the two variables are independent (given by *P(r)P(s)* with entropy equal to *H(R) + H(S)*) and the true observed joint distribution, *P(r,s)* (with entropy *H(R,S)*). This shows that for two independent variables the mutual information between them is equal to zero and illustrates that information measures how far responses and stimuli are from independence.

A common measure of difference between probability distributions is the Kullback-Leibler (KL) divergence (Kullback and Leibler, 1951).

$$D_{KL}(P\|Q) = \sum_x P(x) \log_2 \frac{P(x)}{Q(x)} \tag{4}$$

Note that this is not a true "distance" metric since it is not symmetric – the KL divergence between *p* and *q* is different to that between *q* and *p*. As can be seen from the third expression of Eq. (2), the mutual information is just the KL divergence between the joint distribution and the independent model formed as the product of the marginal distributions:

$$\begin{aligned} I(R;S) &= D_{KL}(P(r,s)\|P(r)P(s)) \\ &= \sum_{rs} P(r,s) \log_2 \frac{P(r,s)}{P(r)P(s)} \end{aligned} \tag{5}$$

In the above *S* and *R*, represent discrete random variables where one is viewed as a stimulus, and the other is a recorded normal response – but these could of course be any two discrete random variables (for example, a variable representing wildtype vs a genetic manipulation, behavioural responses, or other intrinsic signals), and we will see shortly that information and entropy can be easily generalised to continuous signals.

A natural question is "does mutual information correspond to a measure of discriminability?" If by discriminability one means the measure d-prime (Green and Swets, 1966), the answer is in general "no". However, in the specific case where we are measuring the transmission of information about one of two equi-probable stimuli (or equivalently, presence/absence of a stimulus), mutual information has such an interpretation. This can be seen by reaching for another "distance-like measure" – this time, one that is symmetric, the Jensen-Shannon (JS) Divergence, a symmetrized version of the KL Divergence (Lin, 1991; Fuglede and Topsoe, 2004):

$$D_{JS}(P\|Q) = \frac{1}{2} D_{KL}\left(P \middle\| \frac{P+Q}{2}\right) + \frac{1}{2} D_{KL}\left(Q \middle\| \frac{P+Q}{2}\right) \tag{6}$$

It can be fairly easily seen that in the case where we have only two stimuli, $s_1$ and $s_2$, the mutual information *I(R;S)* can be written

$$I(R;S) = D_{JS}(P(r|s_1)\|P(r|s_2)) \tag{7}$$

Thus the mutual information can be considered to measure how far apart (how discriminable) the distributions of responses are, given the two stimuli. Note that while mutual information generalises to multiple stimuli, it is not entirely clear that the concept of discriminability does, in a useful way.

In summary, the mutual information is a measure of how strongly two variables are related, similar in spirit to correlation but with some specific advantages. Firstly, it is a completely general measure; it places no assumptions or models on the variables under consideration and is sensitive to any possible relationships, including non-linear effects and effects in high order statistics of the distributions. Second, it has meaningful units allowing direct comparison across different experiments and even with behavioural performance. Finally, it permits several nice interpretations related to its calculation as a single trial property and involving its relationship to decoding performance.

## Other information theoretic quantities

Similar procedures to those used to estimate entropy and mutual information can be used to estimate a number of other information theoretic quantities. We mention several such quantities here for completeness.

### Multi-information

Note that the mutual information between a pair of random variables naturally generalises to the concept of the multivariate mutual information, or *multi-information*: the mutual information between *n* variables,

$$I(X_1;...;X_n) = \sum_{x_1...x_n} P(x_1,...,x_n) \log_2 \frac{P(x_1,...,x_n)}{\prod_i P(x_i)}$$
$$= I(X_1;...;X_{n-1}) - I(X_1;...;X_{n-1} | X_n)$$
(8)

Note that multi-information, unlike standard mutual information, can take negative values. Multi-information has found use in neuroscience in the study of patterns of activity in neural ensembles (Schneidman et al., 2006). Note that this is not the only way to generalize mutual information beyond two variables, and a related quantity, the interaction information, can also be defined (McGill, 1954; Bell, 2003; Jakulin and Bratko, 2003).

### Conditional mutual information

The conditional mutual information is the expected value of the mutual information between two variables given a third (Cover and Thomas, 1991),

$$I(X;Y|Z) = H(X|Z) - H(X|Y,Z)$$
(9)

This quantifies the relationship between variables X and Y, while controlling for the influence of Z. In neuroscience, this can be useful for example to investigate the coding of correlated stimulus features (Ince et al., 2012). Consider two correlated stimulus features $S_1$ and $S_2$. If it is found that I(R;S1) > 0, this could be because the response is modulated by feature S1, but it might be that the response is modulated by feature S2 and the relationship between response and S1 follows from the correlation between the features. Considering I(R; S1| S2) can resolve this situation and reveal if the feature S1 is truly represented.

### Entropy and information rates

Most biological systems function not as discrete realisations from a static process (like the roll of a die), but rather operate continuously as time-varying dynamic processes. This brings to mind the notion of the *rate* at which a source generates information – e.g. if we were tossing a coin once per second, and the outcomes of the coin tosses were independent, then we would be generating information at a rate of 1 bit/sec. In general, the *entropy rate* of a stochastic process is defined as

$$h(X) = \lim_{n \to \infty} \frac{1}{n} H(X_1, X_2, ..., X_n) .$$
(10)

By extension, the *mutual information rate* is

$$i(R;S) = \lim_{n \to \infty} \frac{1}{n} I(r_1, r_2, ..., r_n; S)$$
(11)

with units of bits/sec. Asymptotic entropy and information rates can therefore be estimated indirectly by calculating information for sufficiently long sequences of time bins. If the calculation is repeated for smaller and smaller time bins it is possible to extrapolate the resulting discrete information value to the instantaneous limit, making the rate calculation independent of both sequence length and bin width (Strong et al., 1998). It is also possible to calculate them directly using a Bayesian probabilistic model (Kennel et al., 2005; Shlens et al., 2007).

## Information theoretic quantities for continuous variables

The quantities above are all defined on random variables taking discrete values. However, entropy and other information quantities can also be defined on continuous spaces. In the case of entropy, replacing summation by integration yields the *differential entropy*:

$$h_{\text{diff}}(X) = \int_X p(x) \log_2 p(x) \tag{12}$$

Other information theoretic quantities can be defined analogously, in general by replacing sums over the discrete spaces with integrals over the continuous spaces. Note that differential entropy does not entirely generalize the properties of the discrete Shannon entropy, and thus this quantity has not been widely used in neuroscience, and is therefore beyond the scope of this article. In contrast, the Kullback-Leibler divergence and mutual information both generalize in a straightforward manner to continuous spaces. For instance, the mutual information between two random variables with joint density *p(x,y)* is

$$I(X;Y) = \int p(x,y) \log_2 \frac{p(x,y)}{p(x)p(y)} dxdy \ . \tag{13}$$

## Acknowledgements

Research supported by the SI-CODE (FET-Open, FP7-284533) project and by the ABC and NETT (People Programme Marie Curie Actions PITN-GA-2011-290011 and PITN-GA-2011-289146) projects of the European Union's Seventh Framework Programme FP7 2007-2013.